\documentclass[12pt]{JHEP3}
\usepackage{epsfig}
\usepackage{color}
\epsfclipon
\usepackage{multicol}
\usepackage{epsfig,bm}
\usepackage{amssymb,amsmath}

\newcommand{\roughly}[1]{\mathrel{\raise.3ex\hbox{$#1$\kern-0.85em
\lower1ex\hbox{$\sim$}}}}

\def\be{\begin{equation}}
\def\ee{\end{equation}}
\def\ba{\begin{eqnarray}}
\def\ea{\end{eqnarray}}

\title{Eternal inflation and a thermodynamic treatment of Einstein's equations}

\author{Jos\'e Tom\'as G\'alvez Ghersi${}^{1,2}$, Ghazal Geshnizjani${}^{2,3}$, Federico Piazza${}^{4}$, Sarah Shandera${}^{2}$\\
$^1$ Facultad de Ciencias, Universidad Nacional de Ingenier\'ia, Lima, Per\'u;\\
$^2$ Perimeter Institute for Theoretical Physics,
  Waterloo, Ontario, Canada;\\
$^3$Department of Physics, SUNY at Buffalo, Buffalo, NY 14260, United States of America;\\
$^4$ PCCP and APC, CNRS (UMR7164), Universit\'e Denis Diderot Paris 7, Batiment Condorcet,  10 rue Alice Domon et L{\'e}onie Duquet, 75205 Paris, France
 }

\date{}

\abstract{In pursuing the intriguing resemblance of the Einstein equations to thermodynamic equations, most sharply seen in systems possessing horizons, we suggest that eternal inflation of the stochastic type may be a fruitful phenomenon to explore. We develop a thermodynamic first law for quasi-de Sitter space, valid on the horizon of a single observer's Hubble patch and explore consistancy with previous proposals for horizons of various types in dynamic and static situations. We use this framework to demonstrate that for the local observer fluctuations of the type necessary for stochastic eternal inflation fall within the regime where the thermodynamic approach is believed to apply. This scenario is interesting because of suggestive parallels with black hole evaporation.}

\preprint{}

\keywords{Effective Field Theory, Cosmology, Inflation, Thermodynamics}

\begin{document}

\section{Introduction}
Black hole thermodynamics~\cite{Bekenstein:1973ur,Bardeen:1973gs}, from the most utilitarian perspective, provides a practical shortcut for describing the black hole evaporation process. In order to predict what happens to an isolated black hole, one should in principle work out the renormalized energy-momentum tensor of the known quantum fields on a black hole solution and plug it back into the Einstein equations. This is known to be a very complicated calculation, involving non-local expressions and, in four dimensions, semiclassical equations of fourth order. Two-dimensional examples of the precise calculation are well studied, largely following the work of Callan, Giddings, Harvey and Strominger~\cite{Callan:1992rs}, but the four dimensional case is so far much less tractable. However, several interesting features of black hole evaporation can be understood from the thermodynamic formulation alone, and in this respect it offers a great simplification~\cite{Page:1976df}. The rate of emission of a black hole is that of a black body of temperature $\mathcal{T}=1/8\pi M$ with grey body factors obtainable in perturbation theory for each species. We can thus sketch the geometry of an evaporating black hole as an approximate Schwarzschild solution with adiabatically varying mass ${\dot M}\sim - 1/M^{2}$ and easily estimate the evaporation time $t_{\rm ev}\sim G^2M^3\sim RS$, where $G$ is Newton's constant, $R$ is the radius of the black hole and $S$ is its entropy. 

What are the limits of applicability of the above description? Naively, the semi-classical treatment is accurate enough for macroscopic black holes whose mass is large in units of the Planck mass, $M \gg M_{\rm Pl}$, and on scales longer than the Planck scale. That is, the scenario seems likely to be under control as long as the back-reaction of the quantized fields is small compared to the typical curvature at the horizon. However, several arguments, nicely summarized in~\cite{ArkaniHamed:2007ky}, indicate that the naive perturbative low-energy approach is only valid on scales parametrically longer than the Planck scale and set by $t_{\rm ev}$; this is a statement of the black hole information puzzle.

It is tempting to speculate that the thermodynamic story can be extended to systems beyond black holes, where it might similarly capture aspects of quantum gravity independent of the microphysics at work on very small scales. Some authors have gone even further and suggested the ambitious program of deriving the Einstein equations from thermodynamics. Jacobson extended the first law of black hole thermodynamics to any local Rindler horizon and showed that $dQ=\mathcal{T}dS$, with suitable definitions of those quantities, contains the same information as the Einstein equation~\cite{Jacobson:1995ab} (see also~\cite{Padmanabhan:2003gd, Padmanabhan:2009vy, Kolekar:2010dm}). Reference \cite{Kothawala:2010bf} contrasts some of these approaches. Hayward~\cite{Hayward:1997jp} and more recently Cai and collaborators have considered spherically symmetric spaces, including FRW  ~\cite{Cai:2005ra, Cai:2010hk, Cai:2010sz}. Multiple fluids in a cosmological setting have been considered in \cite{Jamil:2009eb}. Frolov and Kofman~\cite{Frolov:2002va} considered a quasi-de Sitter space and showed that the evolution of the (perturbed) apparent horizon can also be reformulated as the first law of thermodynamics. Inspired by more recent speculations by Verlinde~\cite{Verlinde:2010hp} and subsequent investigations (e.g.~\cite{Nicolini:2010nb}) it has been shown~\cite{Piazza:2010hz} that a thermodynamic relation can also be applied on a time-like screen, once a few geometrical constraints are met. At the current level, most of the attempts to derive the Einstein equations in fact invoke some knowledge from them (for example in starting with a metric that is a vacuum solution or in choosing a definition of mass or energy flux justified by relating the stress-energy tensor to geometry) and so it is difficult to judge the success of that approach so far. 

From a more conservative perspective, the first step in exploring the thermodynamic framework is to look for cases where geometric quantities take properties similar to those we associate with temperature and entropy. Since the evaporation is the key feature of interest for the black hole case, it may be particularly interesting to consider thermodynamic descriptions of other systems where the back reaction of the quantized fields plays a relevant role. Beyond the computationally practical reason highlighted in the first paragraph, such systems may provide a means to search for and clarify puzzles similar to the black hole information story. Eternal inflation of the stochastic type is a well-known example of such a system and so it is worthwhile to consider whether a thermodynamic treatment can be applied there. 

Our work will be similar to the early black hole work, rather than Jacobson's more general approach: the analysis uses a specific class of solutions (spherically symmetric, assuming adiabatic evolution). We will not try to derive the Einstein equations but rather understand whether a thermodynamic interpretation can apply in the stochastic eternal regime.

For a slow-roll inflation scenario we can exploit the nearness to de Sitter and assume that the temperature associated with the Hubble horizon is $\mathcal{T}\sim H_0$, where $H_0$ is the Hubble parameter at the time of interest. For a spectator scalar field in an exact de Sitter background, one can calculate the contribution from the fluctuations of the field to the background renormalized energy momentum tensor, e.g. $\langle T^{\mu}_{\mu}\rangle_{\rm ren} \sim H_0^4 \sim \mathcal{T}^4$ \cite{Birrell:1982ix}. This is always  a negligible contribution to the total energy density $\rho_{\rm tot} \sim H_0^2 M_{\rm Pl}^2$. In scenarios close to de Sitter considering a single Hubble patch this calculation is expected to hold qualitatively but it is not the only effect of the fluctuations in the eternally inflating regime. There the change in field position due to the fluctuations is as large as the classical rolling of the field down the potential in a Hubble time.  The fluctuations qualitatively change the evolution of the system even though the Hubble parameter itself may (locally) remain nearly constant. In practice one takes the {\it classical} position of the field on the potential to be adjusted by the combination of classical rolling and quantum fluctuations, where this adjustment is homogeneous on a Hubble scale and is performed in steps of order the Hubble time. 

Intuitively it seems reasonable to expect that the phenomena of stochastic eternal inflation, like black hole evaporation, does not depend on a precise formulation of quantum gravity. In order to better define the domain of validity of effective field theory in this context we might try to explicitly suppress the quantum gravity effects (which in the black hole evaporation case are responsible for the recovery of information) and then ask about the validity of our description. The simplest decoupling limit, $M_p\rightarrow\infty$, is too trivial as it makes gravity oblivious of the matter fields' dynamics and kills the scalar fluctuations that are responsible for the chaotic inflationary regime. But because there is an extra parameter in inflation that controls the departure from equilibrium (the slow-roll parameter $\epsilon\equiv-{\dot{H}\over H^2}$), we might hope to decouple UV gravitational effects while still maintaining a dynamical system. Naively, together with sending $M_{\rm Pl}\rightarrow \infty$, we want to send the combination $H^2/\epsilon$ to infinity in such a way that the amplitude of curvature perturbations 
\be
\langle\zeta^2\rangle \sim \frac{H^2}{\epsilon M_{\rm P}^2}
\ee
remains constant (and such that we return neither to the flat space limit, nor the limit where the horizon is too small for quantum modes to created). Tensor modes and graviton loops will be suppressed, and we can conclude that any remaining puzzles are related to IR effects. Although we leave a rigorous analysis of this decoupling limit for further work, it seems very reasonable to expect that this limit or some variant is possible. If the stochastic eternal inflation scenario has a description at sub-Planckian scales it can be a useful test case for the thermodynamic story, analogous to the evaporating black hole.

To that end, we will follow up on a relatively conservative extension of the thermodynamic ideas as considered by Frolov and Kofman~\cite{Frolov:2002va}, where we allow a quasi de Sitter space and small metric fluctuations, requiring spherical symmetry for simplicity. We will first show that one can modify their approach to include contributions from pressure to the first law and that it then yields a temperature consistent with Hawking temperature. We demonstrate that, regardless of what the deep implications may be, the thermodynamic relationship as proposed holds on the horizon of a single observer in near-equilibrium systems during both slow-roll and stochastic eternal inflation. For slow-roll inflation, Frolov and Kofman gave a treatment that included the first order fluctuations of the metric and inflaton field. However, they left the eternal inflating regime out on two grounds: first, that increasing Hubble parameter is incompatible with the classical Einstein equations; second, that the large variation in the Hubble parameter on super-horizon scales indicates that an adiabatic approach is not sufficient. However, the first objection is no obstacle in the black hole evaporation scenario, where the back-reaction of quantum fluctuations also does not behave as a classical matter source. The second objection is removed if we take a local perspective and follow a single Hubble patch: for a local observer the eternal inflation regime is (at least for a short time, and in the slow-roll scenario) perturbative. 

The plan of the paper is as follows.
In the next section we review the definition of the apparent horizon. In Section 3 we set up the first law of thermodynamics in the quasi-de Sitter context and evaluate that equation for slow-roll inflation. In Section 4 we include fluctuations. In Section 5 we demonstrate that the thermodynamic approach finds the same success and difficulty in the eternal inflating regime as the standard general relativistic approach. In the final section we discuss some implications of this result and conclude.

\section{Defining the horizon} \label{defhorizon}
The standard ingredient for applying the thermodynamic relation $dQ =\mathcal{T}dS=dE+PdV$ to spacetime is the presence of an event horizon, which hides information and can thus be associated with an entropy $S$. 
In this context the horizon is inevitably global and observer-dependent; it requires the knowledge of the entire spacetime and that of the observer's world-line in order to be defined. For practical purposes it is useful to look for a different type of horizon, and deal with quantities that can be locally characterized. A review of these issues can be found in ~\cite{Padmanabhan:2003gd}.

In the literature different types of surfaces have been proposed that can be compatible with the notion of entropy. 
In the vicinity of any spacetime point there exist congruences of null rays (with locally vanishing expansion) that form a local Rindler horizon for a class of accelerating observers. Jacobson~\cite{Jacobson:1995ab} applies the first law of thermodynamics to such null rays. He considers the variation of area spanned by the congruence of null-rays and relates it with the flux of matter through the Rindler horizon. Since the congruence is initially non expanding, $dS$ is given by integrating the logarithmic derivative of the expansion which is related to Ricci tensor through the Raychaudhuri equation. In addition, Jacobson's association of this surface with entropy is based on the notion of entanglement entropy.

Frolov and Kofman~\cite{Frolov:2002va}, on the other hand, focus their attention on the \emph{apparent horizon} in spherically symmetric spaces. Their approach is slightly different because they follow the apparent horizon (which is generally a non-null surface) and define $dS$ through the variation of its area. The definition of the apparent horizon relies on the existence of a marginally trapped surface: a two-dimensional space-like surface orthogonal to a congruence of null rays with zero expansion. A marginally trapped surface that is also \emph{closed} generally indicates the presence of strong gravitational effects. The set of marginally trapped surfaces forms an apparent horizon. We will follow Frolov and Kofman and use this surface to define a thermodynamic analogy applicable to inflationary spacetimes (this is also similar to what is proposed for dynamical black holes, e.g.~\cite{Hayward:1994bu, Ashtekar:1999yj, Ashtekar:2000hw, Ashtekar:2004cn}). The association of entropy proportional to the area of this surface comes by analogy with the black hole (for de Sitter, the Gibbons-Hawking entropy \cite{Gibbons:1977mu}). Currently, the most compelling direction to explain all the area/entropy laws may be the notion of entanglement entropy, but no exact calculation has proven this.

\begin{figure}[htb]
\centering
\includegraphics[scale=1]{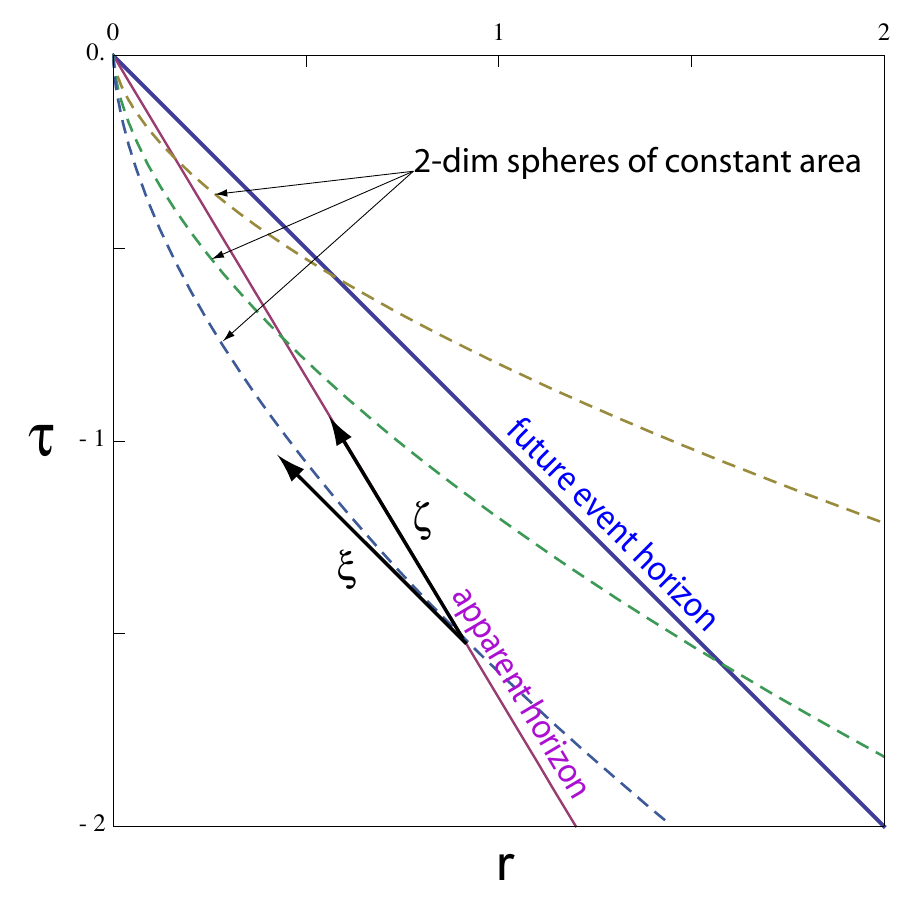}
\centering
\caption{We plot some relevant hypersurfaces of an inflationary spacetime. For illustrative purposes we take $\epsilon= 0.4$. The axes represent the comoving radius $r$ and conformal time $\tau$; radial light rays are lines at 45 degrees. One of them is the future event horizon (EH) for the observer sitting at $r=0$, defined as the past light cone of the observer at future infinity~\protect\cite{Bousso:2004tv}. What happens outside the EH has no influence on the observer. We also plot the apparent horizon (AH), defined by $r = (a H)^{-1}$. The dashed curves are examples of 2-dimensional spheres of constant area. When these surfaces intersect the apparent horizon their tangents are at 45 degrees with respect to the coordinate axes. In other words, the AH is the place where incoming light rays have (instantaneously) zero expansion. The vectors $\zeta$ and $\xi$, on the AH, are parallel to the AH and to the null direction respectively. As summarized in the text, Jacobson's analysis~\protect\cite{Jacobson:1995ab} applies to a null surface, relating the flux of matter through it with the expansion of the null generators (i.e. along $\xi$ in this context). Here, as in~\protect\cite{Frolov:2002va, Cai:2005ra}, we consider the system contained within the AH i.e. we study the matter flux through the AH and relate it with the expansion along the $\zeta$ direction.
The apparent horizon, event horizon, and the surface $\rho=$ const.$=H^{-1}$ all coincide in the de Sitter limit $\epsilon=0$.}
\label{fig:STdiagram}
\end{figure}

In an expanding cosmology, in order to find a closed marginally (anti-) trapped surface, one has to consider larger and larger spheres, until the expansion of the Universe counterbalances the tendency of the ingoing light rays to ``converge". More explicitly, consider a flat FRW metric
\begin{equation}\label{metric}
ds^2 = -dt^2 + a^2(t)(dr^2 + r^2 d\Omega).
\end{equation}
Radial light rays satisfy the equation $dr = \pm dt/a$, where a plus sign holds for outgoing and a minus for ingoing rays respectively. A beam of light rays leaving a sphere of constant $r$ describes two spheres (``$+$" and ``$-$") of areas $A_\pm(t) = 4\pi a^2(t) r_{\pm}^2(t)$, where $r_{\pm}(t)$ are the two solutions for the radial null rays. The derivative of the area along the two beams,
\begin{equation} 
\label{dAsimplest}
\frac{dA_\pm}{dt} = 8 \pi a^2(t) r_{\pm}^2(t)\left(H \pm \frac{1}{ra}\right),
\end{equation}
has a familiar behavior: at small radii ($r a < H^{-1}$) the light rays going inward tend to contract, the ones going outward  tend to expand. However, sufficiently large spheres ``feel" the expansion of the Universe. For $r > ( a H)^{-1}$ both classes of light rays are expanding and the two-dimensional surface is called (anti-)trapped. At any time, the two-dimensional surface with $r = (a H)^{-1}$ is thus marginally trapped and is referred to as apparent horizon. The union of all the marginally trapped surfaces, i.e. the three dimensional surface, $\Sigma:r(t) = (a(t) H(t))^{-1}$, is called a trapping horizon.

Note that a trapping horizon is generally not a null surface; for $\epsilon>0$ it is time-like. For evolving black holes, a great deal of work has gone in to defining local notions of horizons, especially by Ashtekar and collaborators, but this has focused on null and space-like surfaces \cite{Ashtekar:1999yj, Ashtekar:2004cn} which are relevant for stationary and accreting black holes. Interestingly, it is the third case (time-like surfaces) that is most needed for classically evolving quasi-de Sitter space. For more details about trapping horizons we refer readers to Appendix \ref{generaltrappinghorizon}. As is common in literature, in what follows we will refer to both two dimensional apparent horizon and three dimensional trapping horizon, $\Sigma$, as the apparent horizon.

Frolov and Kofman~\cite{Frolov:2002va} associate the entropy $S = A/4G$ to the area $A$, corresponding to two dimensional apparent horizon at anytime.  
For an inflationary FRW universe with a small and constant slow roll parameter $\epsilon\equiv-{\dot{H}\over H^2}$, it is easy to show that the ratio of the area of the apparent horizon to that of the future event horizon is close to one (see Figure \ref{fig:STdiagram}) 
\begin{equation}
{A_{app}\over A_{eve}}=1-2\epsilon\, .
\end{equation}

Since we will eventually consider perturbations of the metric \eqref{metric} it is useful to work with a more general metric ansatz. We will maintain spherical symmetry of the background in order to make plausibly well defined identifications between geometric and thermodynamic quantities. For this purpose consider the metric
\begin{equation} \label{sphmetric}
ds^2 = g_{ab} dx^a dx^b + \rho^2(t,r) d\Omega\, ,
\end{equation}
where $a,b = 0,1$ label the two coordinates $t$ and $r$, and $\rho$ is a generic function of the latter. We will use greek indices $\alpha$, $\beta$ for four-dimensional quantities. In our notation $\nabla$ represents the covariant derivative in $1+1$ dimensions (with respect to $g_{ab}$) and we use the symbol ($;$) to indicate the covariant derivative in $3+1$ dimensions.

The variation of the area of a surface along a trajectory $x^a(\lambda)$  is given by $dA=k^a \nabla_a A d\lambda$, where $k^a={d x^a(\lambda) \over d \lambda}$ and $\lambda$ represents displacement along the trajectory that can be taken to be the time coordinate for time-like or null trajectories. By generalizing the argument given in Eq.(\ref{dAsimplest}) we find the trapped surface for a spherically symmetric metric by considering
\begin{equation}
\frac{dA}{dt} = 8 \pi \rho {dx^a(t)\over dt} \nabla_a \rho = 0,
\end{equation}
where $dx^a$  is null. Since $\nabla_a \rho$ is also a temporal-radial vector, this equation implies that $\nabla_a\rho$ should be a null vector. The condition for a marginally trapped surface in the spherically symmetric case is then:
 \begin{equation}
 (\nabla \rho)^2 = 0
 \end{equation}
Setting the position of apparent horizon by $f\equiv (\nabla \rho)^2 = 0$, $\nabla_a f$ is normal to apparent horizon, while   
 \be \label{defzeta}
 \zeta^a =\varepsilon^{ab}\nabla_b f
 \ee
may be used to represent the (non-angular) tangential vector along this surface. Here $\varepsilon^{ab}$ is the two-dimensional $(t,r)$ Levi-Civita tensor and $\zeta^\alpha$ for indices other than $(t, r)$ are zero. 

For cosmological scenarios $\zeta^{\alpha}$ is time-like, while for exact de Sitter space, $\zeta^{\alpha}$ becomes a null vector and so is parallel and normal to the usual time-like Killing vector at the de Sitter event horizon (which is also the apparent horizon).  Now calculating $k^a$ along the non-angular generators of the surface, since both $\zeta^a(\lambda)$ and $k^a(\lambda)$ are tangential to this trajectory they must be proportional for all values of $\lambda$. For simplicity we assume $\zeta^a=B(\lambda) k^a$ and in the end $B(\lambda)$ will be irrelevant for our analysis. Therefore, the variation of entropy is related to the variation of the area of the apparent horizon by:
 \begin{equation}\label{entropy}
 dS=( B^{-1}d\lambda) \zeta^a \nabla_a \left ({A\over 4G}\right) \, .
 \end{equation}
In analogy with the black hole case, this is the entropy associated with things outside the horizon measured by an observer inside. In the next section we will consider appropriate definitions of mass, pressure and temperature to write the first law.

\section{First law of thermodynamics for the apparent horizon}
Here we will obtain a first law applicable during inflation which can also incorporate perturbations for spherically symmetric solutions. This derivation differs slightly from what was proposed in \cite{Frolov:2002va} in a way which makes the quantities identified with temperature, mass, and pressure more comforting but makes no difference in the actual calculation: the inputs into both their calculation and ours are geometric identities and the Einstein equations. The intermediate step of labeling thermodynamic quantities may give us some physical insight, but at this level is nothing more than a labeling. Our definitions are similar to those used for spherically symmetric spaces (without fluctuations) by Hayward \cite{Hayward:1997jp} and more recently Cai et. al. \cite{Cai:2010sz}, although we are careful to distinguish signs and the physical interpretation for de Sitter compared to black holes. In addition we follow how the slow-roll parameter $\epsilon$ controls the departure from equilibrium and the corrections to the static definitions of geometric and thermodynamic quantities. As long as $\epsilon\ll1$ the system is near-equilibrium and deviations from exact de Sitter provide only higher order corrections to the thermodynamic equation we will find. However it may be important to keep track of the $\epsilon$ corrections since the system of real interest here cannot have $\epsilon=0$ (in that case there is only a trivial limit for decoupling gravity and the scenario may lose much of its power as a test case).  

\subsection{Energy flow and surface gravity}
To make the analogy between the Einstein equations and the first law, we need to introduce quantities that will play the role of temperature and energy. Precise, globally well defined notions of mass and surface gravity (temperature) are only possible for static, asymptotically flat spaces, although a great deal of work has been done to define quantities that are useful for astrophysical, evolving black holes. Here we will exploit spherical symmetry in order to define local quantities, and we will ensure that the standard results are recovered for static Schwarzschild and de Sitter spacetimes. 

For a spherically symmetric metric, given in Eq.(\ref{sphmetric}), the relation between the four dimensional Ricci scalar $^{(4)}R$ and that derived with the two dimensional metric $g_{a b}$ is
\begin{equation}
^{(4)}R = R - 4 \frac{\Box \rho}{\rho} + \frac{2}{\rho^2}[1 - (\nabla \rho)^2] \;,
\end{equation}
where $\Box\rho$ is defined from the two-dimensional metric.
By integration over the angular variable we obtain the Einstein Hilbert action for a spherically symmetric space,
\begin{equation}
S_{EH} = 4\pi\int d^2 x \sqrt{-g}\left[\frac{1}{16\pi G}(\rho^2 R + 2 (\nabla \rho)^2 + 2) + \rho^2 {\cal L}_{\rm matter}\right],
\end{equation}
and the $(a~b)$ Einstein equations
\begin{eqnarray} \label{Einsteq}
2\rho^{-1}[g_{ab}\Box \rho- \nabla_b \nabla_a\rho]+\rho^{-2}g_{ab}[(\nabla\rho)^2-1]&=&8\pi G T_{ab}\, .
\end{eqnarray} 

\subsection{Defining $\delta E$}
To associate an appropriate quantity to the change in energy, $\delta E$, we make use of the fact that for spherically symmetric but not necessarily static systems there still exist conserved currents \cite{Hayward:1994bu} based on which one can define the flux transfer across a hypersurface. In particular we will use a definition of $\delta E$ which is related through the Einstein equations to $\delta M$ where $M$ is the Misner-Sharp mass, a well-defined local quantity for spherically symmetric metrics. 

While in time-dependent space-time there is no Killing vector for time translation, Kodama has shown \cite{Kodama:1979vn} (also see \cite{Abreu:2010ru}) that for spherically symmetric metrics there is a divergence free vector $\xi^\alpha$ such that, 
\be \label{kodama}
\xi^a\equiv-\varepsilon^{ab}\nabla_b\rho, 
\ee
and $\xi^\alpha = 0$ for indices other than $(t, r)$. It is easy to show this vector satisfies $\xi^\alpha_{;\alpha}=0$ and is null on the surface of the apparent horizon. In addition Kodama proved that $G^{\alpha\beta}\xi_{\beta}$ also corresponds to a conserved current \cite{Kodama:1979vn}. Applying the Einstein equations to this expression we can define a current: 
\be
\label{current}
j^\alpha \equiv -T^{\alpha\beta}\xi_{\beta}
\ee
which is also covariantly conserved, $j^\alpha;_\alpha=0$, and the minus sign ensures the flow for quasi-de Sitter is future directed. This enables us to calculate the energy flux through the apparent horizon,
\be \label{Ej}
\delta E =  j^\alpha d\Sigma_\alpha. 
\ee
$d\Sigma_\alpha$ denotes the 3-volume element of the apparent horizon and it can be written in terms of  $\zeta^a$:
\ba\label{sigmaf}
d\Sigma_a&=&4\pi\rho^2 k^b\varepsilon_{ab}d\lambda \nonumber \\
&=&(B^{-1}d\lambda) 4\pi \rho^2 \zeta^b\varepsilon_{ab} , 
\ea
while $d\Sigma_\alpha = 0$ for indices other than $(t, r)$. 

Although we will consider the current above to be the definition of $\delta E$, it is important to emphasize how the choice is motivated. The original current of Kodama was geometric, and the current $j^a$ above is related through the Einstein equations to the same well-defined quasi-local quantity for spherically symmetric metrics \cite{Abreu:2010ru}, 
\be \label{jM}
j^a={1\over 4\pi\rho^2}\varepsilon^{ba}\nabla_bM,
\ee
where $M$ is defined by
\be \label{massdef}
M \equiv \frac{\rho}{2 G}[1 - (\nabla \rho)^2]. 
\ee
In the literature $M$ is referred to as Hawking-Israel/Hernandez/ Misner-Sharp mass and a thorough discussion of its properties can be found in \cite{Hayward:1994bu}. Now combining Equations \eqref{Ej}, \eqref{sigmaf}, \eqref{jM} and \eqref{defzeta}, we obtain 
\be
\delta{E}= (B^{-1}d\lambda) \zeta^a\nabla_aM 
\ee 
Taking a derivative of above mass and using the (two-dimensional) Einstein equations, yields that $\delta E$ is also related to stress energy tensor in the following way:
\ba \label{Ma}
\delta{E}&=& \frac{1}{2 G}  (B^{-1}d\lambda)\left[(1 - \nabla \rho^2) \nabla_a \rho - \rho \nabla_a (\nabla \rho)^2 \right] \zeta^a\nonumber\\
&=&(B^{-1}d\lambda)4\pi \rho^2(T_{ab}-Tg_{ab})\nabla^b \rho~ \zeta^a\;.
\ea

\subsection{Defining surface gravity}
The other key ingredient we need is a prescription for the surface gravity, $\kappa$. The standard geometric definition is only applicable in a stationary spacetime in the presence of a Killing horizon and one can propose various strategies to extend it to the dynamical apparent horizon (the marginally trapped surface). In what follows we take the approach proposed by Hayward \cite{Hayward:1997jp} for spherical symmetric metrics which produces expected values for known stationary cases. The new definition can be expressed as,
\be\label{defkdyn1}
\xi^a\nabla_{[a}\xi_{b]}=\kappa \xi_b
\ee
or equivalently as,
\be\label{defkdyn2}
\kappa=\frac{1}{2}|\Box\rho|\;.
\ee
For details on why these expressions agree, even up to the sign, and also how they relate to standard definitions of surface gravity we refer the readers to Appendix \ref{appsurfacegravity}. It is difficult to be rigorous, so we content ourselves here with the fact that this definition gives $\kappa=H$ for static de Sitter, and $\kappa=\frac{1}{4M}$ for Schwarzschild, and that our results won't be sensitive to the distinction between this definition and others that reduce to the correct answer in the static case. As discussed in the appendix, there are several proposals for defining surface gravity for non-static cases. They all agree in the static limit but may differ in corrections proportional to $\epsilon$. Since our thermodynamic expression will already be first order in $\epsilon$, these differences in definition do not affect our basic result.

The crucial element for considering event horizons as thermodynamic objects is Hawking radiation~\cite{Hawking:1974sw, Fulling:1972md, Davies:1974th, Gibbons:1977mu, Unruh:1976db}, which has a thermal spectrum with temperature $\mathcal{T}=\frac{\kappa}{2\pi}$. In exact de Sitter space the temperature of quantum fluctuations is related by the same formula to the surface gravity $\kappa_{dS} = H$ that any comoving observers associates with their de Sitter horizon. Defining temperature for a dynamical horizon is a more challenging issue. In highly dynamical situations the system does not have time to reach an equilibrium and the canonical notion of temperature does not exist. This implies that even though one may still interpret which term represents the heat absorption in a meaningful way, there is no clear way to distinguish between entropy and temperature. Similarly for highly dynamical horizons one does not expect the existence of a quantity that can be directly interpreted as an instantaneous physical temperature of some radiation.
However, if we consider slowly evolving scenarios (small departures from de Sitter) it seems reasonable to associate a temperature to the local geometry at any particular time. Therefore, for quasi-de Sitter space with $\epsilon=-\frac{\dot{H}}{H^2}\ll1$, we expect that any corrections to the temperature be proportional to $\epsilon$, that is $\mathcal{T}=\frac{H}{2\pi}(1+\mathcal{O}(\epsilon))$. One may be able to make more precise statements about the adiabaticity conditions following the work of \cite{2004PhRvL..92a1102B} or \cite{Barcelo:2010xk}.

Note that in the laws of equilibrium thermodynamics it is also usually considered that transitions happen from one nearly time independent state to another one. This implies that in infinitesimal physical processes the change in surface gravity in the term that represents $\mathcal{T}dS$ is negligible the same way that change in the temperature is ignored in standard thermodynamics. Inferring temperature as $\mathcal{T}\equiv {\kappa\over 2\pi}$ using the definition of Equation \eqref{defkdyn2} we obtain
\be
\mathcal{T}={H\over 2\pi}\left(1-{\epsilon\over 2}\right), 
\ee
which agrees with our intuition.

The temperature associated to the apparent horizon by Frolov and Kofman \cite{Frolov:2002va} was different as they interpreted a different geometric quantity as $\kappa$, based on a guessed analogy with the black hole geometry:
\be
\kappa_{FK}=\frac{1}{2\rho}\left[1-(\nabla\rho)^2\right]\;.
\ee
They wrote a geometric equation that looked similar to the first law (with zero pressure) where this quantity, $\kappa_{FK}$, appeared as a constant of proportionality between $dE$ and $dS$. However, this choice evaluates to $H/4\pi$ on the horizon, half of the standard Hawking temperature. We will write a similar geometric equation but identify the temperature as the geometric quantity given in Eq.(\ref{defkdyn2}). The difference between our results and theirs underscores a primary difficulty in the thermodynamic interpretation: there are many mathematically correct combinations of geometric identities and the Einstein equations for inflating space times, but without the additional input (e.g. calculating the Hawking temperature) it is difficult to unambiguously determine which quantities (if any) have truly meaningful thermodynamic interpretations.

\subsection{Einstein equations as the first law}

We now begin from the two-dimensional Einstein equations and work to construct something that resembles the first law. Taking Eq.(\ref{Einsteq}) and subtracting half of its (two dimensional) trace multiplied by $g_{ab}$ we obtain
\be \label{Einsteqsubtrace}
-\nabla_b\nabla_a \rho+\frac{1}{2} g_{ab}\Box \rho =4\pi G\rho(T_{ab}-T^c_cg_{ab})+2\pi G \rho T^c_c g_{ab}\;.
\ee
Contracting both sides with $\rho\zeta^a \nabla^b \rho$ and using our definition of $\kappa$ and $M$ above we find
\be \label{roughfirstlaw}
-\zeta^a\nabla_a \left(\frac{A}{4G}\right)\frac{\kappa}{2\pi}=\zeta^a \nabla_a M+\frac{1}{2}T^c_c\zeta^a \nabla_a\mathcal{V}
\ee
where $A$ is the area of the apparent horizon and $\mathcal{V}\equiv 4\pi \rho^3/3$ is the volume it encloses. This equation has been noted by Hayward \cite{Hayward:1997jp} as first law of black-hole dynamics and we see here that it also applies to quasi de-Sitter space-times.
To zeroth order in $\epsilon$, the trace of the stress energy tensor is $T_c^c=2P$, where $P$ is the pressure. Therefore, it looks natural to identify the last term in the last equation as a pressure term, $P \simeq T_c^c/2$. 

In summary, we identify $\frac{\kappa}{2\pi}$ as the temperature associated with the apparent horizon; the mass $M$ and volume $\mathcal{V}$ correspond to the mass an volume {\it within} the horizon. 
We assume that even if we do not have well-defined expressions for the total energy or volume outside the horizon, we still have
\ba
\delta E_{out}&=-&\delta E_{in}\\\nonumber
\delta \mathcal{V}_{out}&=-&\delta \mathcal{V}_{in}
\ea 
For $\delta E_{in}$ in particular, one can check from the signs in Eq.(\ref{current}) and Eq.(\ref{sigmaf}) that both the current and the directed surface element are pointing inwards from the horizon. Then, consistently writing all quantities relevant to the space outside the horizon (analogous to the expressions for the black hole, where the quantities are {\it inside} the horizon) Eq.(\ref{roughfirstlaw}) can suggestively be written as
\ba \label{firstlaw}
\underbrace{\zeta^a\nabla_a\left(\frac{A}{4G}\right)\frac{\kappa}{2\pi}}_{\mbox{$\mathcal{T}\delta S$}}=\underbrace{-\zeta^a \nabla_a M}_{\mbox{$+\delta E_{out}$}}\underbrace{-P\zeta^a\nabla_a\mathcal{V}}_{\mbox{$+P\delta\mathcal{V}_{out}$}}\;. 
\ea
Here to make the analogy more consistent we should have multiplied the equation by the normalization factor, $B^{-1}d\lambda$, but since in the end this factor appears equally in all terms we can ignore it. 

For the unperturbed metric we evaluate this expression for a scalar field source, taking $\delta E_{in}=-\delta E_{out}$ to be defined from Eq.(\ref{current}). We find
\be
\label{zeroOrder}
2M_p^2H^2\epsilon=\dot\phi_0^2\;.
\ee
Interestingly, although the potential $V(\phi_0)$ appears in each of the terms on the right hand side of the thermodynamic equation, all dependence on $V(\phi_0)$ cancels out in arriving at the expression above. The usual derivation of Eq.(\ref{zeroOrder}) in GR uses the continuity equation combined with the $(00)$ Einstein equation (the first Friedmann equation). The thermodynamic version basically inverts that reasoning: in addition to the equation above, if we assume we still have the continuity equation $T^{\alpha}_{\;\;\beta;\alpha}=0$, then the $\beta=0$ equation relates the time derivative of energy density $\rho_E$ to energy density and pressure. Combining that with Eq.(\ref{zeroOrder}) above, we recover the time derivative of the $(00)$ Einstein equation:
\ba
\dot\rho_E&=&-3H(\rho_E+P)\\\nonumber
\frac{d}{dt}\left(\frac{1}{2}\dot\phi_0^2+V(\phi_0)\right)&=&-3H\dot\phi_0^2=6M_p^2H\dot{H}
\ea
So we ``recover" the Einstein equation up to an integration constant, similarly to Jacobson's result \cite{Jacobson:1995ab}. 

The result in Eq.(\ref{zeroOrder}) is the same as found by Frolov and Kofman. To match their starting expression, however, consider the vector $\xi^{a}$ defined in Eq.(\ref{kodama}), the radial generator of $\rho=constant$ surfaces. For exact de Sitter, $\xi^{a}$ becomes null and parallel to $\zeta^{a}$. Using Eq.(\ref{Ma}) it can be shown that, on the horizon, $\zeta^a \nabla_a M=-4\pi \rho \xi^a T_{ab}\nabla^b \rho$. Then we can write Eq.(\ref{roughfirstlaw}) as
\be
\zeta^a \nabla_a \left(\frac{A}{4G}\right) \frac{\Box \rho}{4\pi}= - 4 \pi \rho \xi^a T_{ab} \nabla^b \rho + \frac{1}{2}T^c_c \zeta^a  \nabla_a\mathcal{V}\;.
\ee
Finally, substituting for $T^c_c$ using the trace of Eq.(\ref{Einsteq}) we find the simple form previously derived in \cite{Frolov:2002va},
\be \label{xiformat}
-\zeta^a \nabla_a A=16\pi G AT_a^b\xi^a \nabla_b \rho
\ee
which interestingly relates the change of area of the apparent horizon with the flux of energy passing through the $\rho=constant$ surface. 

\section{Including Perturbations}
To examine what happens in the context of standard slow-roll inflation or eternal inflation of the stochastic type we need to include fluctuations in the analysis above. Since we have so far used only the Einstein equations and geometric identities, we expect the ``first law" in Eq.(\ref{firstlaw}) to hold whenever the Einstein equations do. We will explicitly demonstrate that here, including fluctuations of type required for stochastic eternal inflation. 

\subsection{Perturbing the spherically symmetric metric}
Since we have used spherical symmetry to establish a framework for our calculations, it is convenient to discuss metric perturbations in a format that shows explicitly when that symmetry is preserved. The full theory of metric perturbations about a spherically symmetric background is well-studied because of its application to black holes, and we will follow that literature (concisely reviewed for example in \cite{Martel:2005ir}). 

All metric fluctuations, which need not themselves preserve the spherical symmetry, can be written as expansions in spherical harmonics multiplied by functions of $r$ and $t$. However, to illustrate our main point we don't need to consider all possible modes. To see why write the the source term, fluctuations in the scalar field, as an expansion in spherical harmonics \cite{Frolov:2002va}:
\ba
\delta\phi(t,{\bf x}) &=&\sum_{\ell m}\int dk\left(\hat{a}_{k\ell m}\phi_{k\ell m}+\hat{a}^{\dag}_{k\ell m}\phi^{*}_{k\ell m}\right)\\\nonumber
\phi_{k\ell m}&=&\phi_k(t)2kj_{\ell}(kr)Y_{\ell m}(\theta, \phi)\\\nonumber
\phi_k(t)&=&\frac{\sqrt{\pi}}{2}H\tau^{3/2}H^{(1)}(k\tau)
\ea
where $j_{\ell}(kr)$ are spherical Bessel functions, $Y_{\ell m}(\theta,\phi)$ are spherical harmonics, $H^{(1)}(k\tau)$ are Hankel functions of the first kind, $\tau\sim-1/(aH)$ is conformal time and the creation and annihilation operators $\hat{a}_{k\ell m}$, $\hat{a}^{\dag}_{k\ell m}$ act on the Bunch-Davies vacuum. In other words, this is the usual solution but with momentum expanded in spherical harmonics rather than Cartesian modes. In particular, the modes still have the behavior of oscillating while well inside the horizon ($k\tau\gg1$) and freezing out on scales of the horizon size. For $\ell\neq0$ modes, $j_{\ell}(kr)\rightarrow0$ as $kr=k\tau\rightarrow0$ and even on horizon scales, $r\sim (aH)^{-1}$, these modes are less important. We need only consider the $\ell=0$ mode which dominates there. 

This makes life simpler in terms of the metric perturbations, as we can consider only the $\ell=0$ mode perturbations (which preserve the spherical symmetry). For a generic spherically symmetric metric, these are \cite{Martel:2005ir}
\ba
ds^2&=&\left(g_{ij}+h_{ij}Y^{00}\right)dx^idx^j\\\nonumber
&&+\rho^2(x^i)\gamma_{\alpha\beta}(x^{\alpha})(1+Y^{00}K)dx^{\alpha}dx^{\beta}\;.
\ea
We can write this in a form that is more familiar by defining
\ba
-2\Phi&\equiv& h_{tt}Y^{00}\\\nonumber
-aB_{,r}&\equiv &h_{rt}Y^{00}\\\nonumber
-2a^2\Psi_1&\equiv&h_{rr}Y^{00}\\\nonumber
-2a^2r^2\Psi_2&\equiv&r^2Y^{00}K
\ea
and applying this to the usual FRW metric:
\be
\label{eq:SpherCNgauge}
ds^2=-(1+2\Phi)dt^2-2aB_{,r}drdt+a^2(1-2\Psi_1)dr^2+a^2r^2(1-2\Psi_2)(d\theta^2+\sin^2\theta d\phi^2)\;.
\ee
A generic gauge transformation on this mode takes $x_{\alpha}\rightarrow \tilde{x}_{\alpha}=x_{\alpha}+\Xi_{\alpha}$ with $\Xi_{\alpha}=Y^{00}(\xi_0,\xi_1,0,0)$. Our perturbations transform as
\ba
\Phi\rightarrow\Phi+\dot{\xi}_0\;,&\;\;\;&B_{,r}\rightarrow B_{,r}-a^{-1}(\dot{\xi}_1-\partial_r\xi_0)+2\frac{\dot{a}}{a}\xi_1\\\nonumber
\Psi_1\rightarrow\Psi_1-\frac{\dot{a}}{a}\xi_0+a^{-2}\partial_r\xi_1\;,&\;\;\;&\Psi_2\rightarrow \Psi_2+\frac{1}{a^2r}\xi_1-\frac{\dot{a}}{a}\xi_0
\ea
For this limited case, we cannot define gauge-invariant variables, but we can choose a gauge that looks familiar by setting $B_{,r}=0$ and $\Psi_1=\Psi_2$. Then comparing the $G_{11}$ and $G_{22}$ Einstein equations we find $\Phi=\Psi$ and so we have the familiar conformal Newtonian gauge (also used by Frolov and Kofman):
\be \label{apparent-newtonian}
ds^2=-(1+2\Phi)dt^2+(1-2\Phi)a^2(dr^2+r^2d\Omega^2)
\ee
For the perturbed metric the apparent horizon is at
\be
\rho_H=H^{-1}(1+\Phi+H^{-1}\dot{\Phi}-(aH)^{-1}\Phi_{,r})\:.
\ee
The $(0,1)$ Einstein equation gives
\ba
\label{eq:00Eeqn}
(\dot\Phi+H\Phi)_{,r}&=&-4\pi G (-\dot\phi_0\delta\phi)_{,r}\\\nonumber
\dot\Phi+H\Phi&=&4\pi G\dot\phi_0\delta\phi
\ea
and the $(0,0)$ Einstein equation gives (using the previous equation and the equation of motion)
\be
\left(\frac{\nabla^2}{4\pi Ga^2\dot\phi_0^2}+1\right)\Phi=\frac{d}{dt}\left(\frac{\delta\phi}{\dot\phi_0}\right)\;.
\ee
Substituting these expressions into $2\dot\phi\dot{\delta\phi}=\dot\phi^2(\delta\phi/\dot\phi)^{\cdot}+(\dot\phi\delta\phi)^{\cdot}$ and adding the top line of Eq.(\ref{eq:00Eeqn}) gives an expression we will be able to compare to the `thermodynamic' equation:
\be
\label{EinsteinCompare}
\ddot\Phi-\frac{2}{a}\dot\Phi_{,r}+\frac{1}{a^2}\Phi_{,rr}+H\dot\Phi+\frac{2}{a^2}\left(\frac{1}{r}-\dot{a}\right)\Phi_{,r}=8\pi G\dot\phi\left(\dot{\delta\phi}-\frac{1}{a}\delta\phi_{,r}\right)\;.
\ee

\subsection{Comparing the `thermodynamic' equation}
Now we will compare the result of the Einstein equations above to the `thermodynamic' equation. We again evaluate Eq.(\ref{firstlaw}), including all terms first order in metric perturbations and in scalar field fluctuations. We find (for arbitrary values of $\epsilon$) 
\be
\label{eq:thermoFluct}
\ddot\Phi-2a^{-1}\dot\Phi_{,r}+a^{-2}\Phi_{,rr}+H\dot\Phi=8\pi G \dot\phi_0(\dot{\delta\phi}-a^{-1}\delta\phi_{,r})\, .
\ee
This agrees with the Einstein equation above, Eq.(\ref{EinsteinCompare}), up to a term which vanishes on the horizon. (It also agrees with the final expression found by Frolov and Kofman). Near the original horizon, $rH_0=1$ and the expressions are equal. The position of the horizon has been perturbed by an amount $\Phi$ so that $\Phi<1$ is required. Using the second line of Eq.(\ref{eq:00Eeqn}), the condition $\Phi<1$ implies $H\sqrt{\epsilon}<M_p$, which is obviously satisfied during inflation. 

\section{Eternal Inflation}
Despite the serious difficulties presented by the global picture of an eternally inflating multi-verse, the local picture of a Hubble patch in which the inflaton field fluctuates up the potential has a well-controlled perturbative description in the sense that metric and field fluctuations are small. In some gauges, for example a slicing where we choose to follow constant $\phi$ hypersurfaces, the fluctuations may not look small, but this is no obstacle to choosing a different slicing in which to follow the story \cite{Creminelli:2008es,Leblond:2008gg}.

To illustrate this, we first verify that in the gauge above the metric fluctuations are small even in the eternally inflating regime. The condition for eternal inflation is that quantum fluctuations, which may move the inflaton field up or down along the potential, are as large on average as the classical distance the field would move down the potential in a Hubble time. In other words: 
\ba
\label{SEIcondition}
\frac{H(\delta\phi)}{\dot\phi_0}&\sim&1\\\nonumber
\sqrt{2\epsilon}&\sim&\frac{H}{M_p}\;.
\ea
Since the curvature perturbation which remains constant outside the horizon has an amplitude
\be
\sqrt{\langle|\zeta|^2\rangle}=\frac{H}{M_p\sqrt{2\epsilon}}
\ee
we see that at the eternal inflation boundary on slices of constant density ($\delta\phi=0$), curvature fluctuations are of order one. However, this does not cause any problem with our analysis in the previous section. There, we saw that our solution for $\Phi$ was small regardless of the condition in Eq.(\ref{SEIcondition}). In other words, we found that $\Phi$ evaluated using the horizon crossing expectation value for $\delta\phi$ was
\be
|\Phi|\sim\frac{H\delta\phi}{|\dot{\phi}_0|}\cdot\frac{\dot{\phi_0}^2}{M_p^2H^2}
\ee
so that even if the first factor is order one, the second factor must be small while we have inflation (and is smaller the flatter the potential). 

Still, something interesting is happening in the eternal regime. Thanks to the quantum fluctuations $H$ can remain nearly constant as required but its derivative, $\dot{H}$, is no longer determined by the classical motion of the scalar field. In the usual GR calculation, we take a linear combination of the $(00)$ and $(11)$ Einstein equations to solve for $\dot{H}$, and it is exactly this linear combination that the `thermodynamic' version of the equation cares about (we've asked for how the horizon area, essentially determined by $H$, is changing). So, in fact the thing that is breaking down in the eternally inflating regime is the separation into background and fluctuation we did in the previous sections, writing separately Eq.(\ref{zeroOrder}) and Eq.(\ref{eq:thermoFluct}). Using the naive scaling that in the eternally inflating regime $\dot{\delta\phi}\sim H\delta\phi\sim\dot\phi_0$, we see by comparing the right hand sides of Eq.(\ref{zeroOrder}) and Eq.(\ref{eq:thermoFluct}) that the source terms for $\dot{H}$ and $\Phi$ are roughly the same size (and there are other terms like $(\dot{\delta\phi})^2$, which we dropped above, that are also of the same size). This problem is the same in both the standard GR treatment of the problem and in the thermodynamic treatment since we are merely looking at a particular combination of the Einstein equations restricted to a particular surface.

In practice, we don't know how to formally treat a series of such steps, and so the stochastic inflation picture \cite{Starobinsky:1986fx} was developed to capture the long-time behavior of the system. This has provided a framework for thinking about a more global picture of eternal inflation, where we ask about statistics of the field on very large scales, coarse-grained over roughly Hubble scales. However, if we wanted to go further with the local picture, the calculation is in some ways similar to what must be done for an evaporating black hole: we want to calculate the back-reaction of a quantum process on the metric. Some difference (and difficulty) comes from the fact that we must consider a system that already has classical evolution ($\epsilon\neq0$) to see the quantum effects for de Sitter, and so the back-reaction calculation is in practice more complex. (On the other hand, there has been some previous work applying the stochastic formulation to black hole evaporation \cite{Hu:2007tq}). However, as with the evaporating black hole, we may not need to actually do that calculation precisely in order to learn something from the system.

Of course there is also a question of the second law, and whether perhaps there is a generalized second law for de Sitter space by analogy with the evaporating black hole. A traditional approach to this problem has been to rely on the fact that for any given Hubble patch the decrease in de Sitter entropy due to stochastically increasing $H$ is small \cite{Bousso:2006ge} since the patch eventually exits eternal inflation back into the slow-roll regime. It would be very interesting to investigate if something further can be said.

\section{Discussion}

In summary, we have revisited the question of thermodynamics and eternal inflation of the stochastic type. We have found that a version of the first law, suggested previously in \cite{Hayward:1997jp}, but largely examined in the context of black holes, seems to be reasonably well-motivated and consistent for quasi-de Sitter space as well. Using this expression, we demonstrated that a local description of eternal inflation of the stochastic type falls within the regime where the thermodynamic picture can be defined. This is not surprising, but is a consequence of the fact that there is a perturbative, local description of stochastic eternal inflation in GR and we have used only the Einstein equations and geometric identities to arrive at the thermodynamic equation. A more non-trivial test of the suggested ``first law" would come from checking the next order corrections in $\epsilon$, which would help determine the best definitions for surface gravity and entropy. It may also be useful to consider if the thermodynamic relations change when one allows perturbative departures from the Einstein Hilbert action (higher curvature terms). For black holes, this has been studied since the 1980's. In the context of Jacobson's work\footnote{We thank the referee for bringing this literature to our attention.}, the investigation began in \cite{Eling:2006aw} and more recent work includes \cite{Brustein:2009hy, Chirco:2009dc, Chirco:2010sw, Padmanabhan:2010rp}.  Related considerations for static, spherically symmetric spacetimes are considered in \cite{Kothawala:2009kc} and FRW spacetimes in \cite{Wu:2009wp}.

However, the implications of even this lowest order result are interesting: both black hole evaporation and stochastic eternal inflation are clues to infra-red puzzles in gravity. There are hints that the same long time scale that sets the black hole evaporation time indicates where an effective field theory description first breaks down \cite{ArkaniHamed:2007ky}, $t_{ev}=RS$. Something similar appears in de Sitter computations (for a recent summary see \cite{Giddings:2010nc} and references therein) where $R$ and $S$ are the radius and entropy of the (quasi) de Sitter space rather than of the black hole. Those hints suggest that, in the regime of dynamical gravity, the global de Sitter solution is just an approximation; more generally, it seems that the validity of semi-classical GR may be limited in the infra-red\footnote{Other research directions, somewhat orthogonal to this discussion, share similar conclusions. 
An infra-red geometric modification of semiclassical GR has been envisioned~\cite{Piazza:2009ek,Piazza:2009bp} such that $\langle T_{\alpha \beta}\rangle$ evaluates the same as in flat space and the non-local terms responsible for the backreaction are absent from the beginning.}, and not only in the UV as naively expected. 
One of the more interesting claims of the thermodynamic approach is that it gives insight into gravity independent of the particular microphysical details of the theory. As discussed in the introduction in some detail, we expect the regime of validity of the thermodynamic treatment to be the same as that of semi-classical gravity. If this idea has depth, stochastic eternal inflation is an ideal laboratory, complementary to the evaporating black hole, in which to test it.

Alternatively, the thermodynamic analogies may be indications that we should not try to quantize gravity in the traditional way. This point of view would be strengthened if some thermodynamic argument could provide testable effects \emph{beyond} Einstein gravity. It is tempting to speculate further and give an example inspired by the present calculation. 
As in all the thermodynamic versions of the Einstein equations derived so far (e.g. ~\cite{Jacobson:1995ab,Frolov:2002va,Padmanabhan:2009vy,Piazza:2010hz}), our first law~\eqref{firstlaw} is oblivious to a cosmological constant term in the energy momentum tensor. This can be checked explicitly by substituting $T_{ab} = - \Lambda g_{a b} $ on the RHS of the equation and noting that, in this case, $P = -\Lambda$. That is, in a cosmological setup, our equation (together with the continuity equation) is equivalent to the \emph{time derivative} of the Friedman equation, and $\Lambda$ is just an integration constant. At present, however, this is no more than a curiosity that appears thanks to our choice of definition of $\delta E$. If gravity were really insensitive to $\Lambda$, this would suggest deviations from our current picture of GR, such that the quantum cosmological constant generated from matter fields is zero, or that the value is dynamically irrelevant. For example there should exist a mechanism that, during a phase transition, adiabatically readjusts to zero the ``zero mode" vacuum energy contribution from the matter fields. Because they ``just" reproduce the Einstein equations, thermodynamic analogies have not revealed a mechanism of this type so far, but this could be yet another direction to explore.

\section*{Acknowledgements}
We thank Niayesh Afshordi, Andrei Frolov, Louis Leblond, Luis Lehner and Rafael Sorkin for many important discussions, especially regarding the well-traveled aspects of this problem. This research was supported by Perimeter Institute for Theoretical Physics. Research at Perimeter Institute is supported by the Government of Canada through Industry Canada and by the Province of Ontario through the Ministry of Research \& Innovation.

\appendix

\section{Surface gravity}\label{appsurfacegravity}
\subsection{Review of surface gravity for a static spherically symmetric space-time}
 Consider a static spherically symmetric space-time with metric \footnote{This is not the most general static spherically symmetric metric but for perfect fluid sources this is sufficient. For more general cases, see the literature on ``dirty black hole'' metric  solutions.}
 
 \be \label{metricstatic}
ds^2=-f(\rho)dt^2+f(\rho)^{-1}d\rho^2+\rho^2d\Omega^2. 
\ee
We now follow~\cite{2004rtmb.book.....P}, keeping track of the signs carefully. If a particle is held stationary at constant $\rho$ its acceleration, $A^\alpha$ can be computed as
\ba
A^\alpha= u_{;\beta}^\alpha u^\beta,
\ea
where $u^{\alpha}=dx^\alpha /d \tau $  is the four velocity along this trajectory and $\tau$ is proper time. One can easily see that $A^0=0$, while $A^\rho={1\over2 }f'(\rho)$. Therefore, the magnitude of the local force required to hold a unit mass at this position (equated to the magnitude of acceleration) is
\ba
A={1\over2 }f^{-1/2}(\rho)|f'(\rho)|.
\ea 
Note that at the horizon $f(\rho_H)=0$ and the force is divergent. However, if this mass is held at that point with a string by an observer at a different radius $\tilde{\rho}$, energy conservation arguments~\cite{Wald:1984rg,2004rtmb.book.....P} show that the force that she exerts on the string is in fact multiplied by a redshift factor, 
\begin{equation}
\tilde{A} = \left(\frac{f(\rho)}{f(\tilde{\rho})}\right)^{1/2} A.
\end{equation}
This force is generally finite as the infinite force at horizon is diluted by an infinite redshift. By choosing $f(\tilde{\rho})=1$ we arrive at the expression for the surface gravity
\be \label{sgfprime}
\kappa={1\over2 }|f'(\rho_H)|.
\ee
Note that  the above definition is intrinsically non local as it relies on the position of the reference observer at $\tilde{\rho}$ for which $f(\tilde{\rho})=1$. The obvious choices for a black hole and de Sitter in static coordinates are to put the observer at $\tilde{\rho} \rightarrow \infty $ and $\tilde{\rho} =  0$ respectively.

One can also make use of time translation Killing vector to define surface gravity.  For a static metric $\xi^\alpha\equiv \partial x^\alpha/\partial t$ is the  time translation Killing vector. Note that  any constant multiple of $\xi^\alpha$ is also a Killing vector. However, $\xi^\alpha$ is conventionally taken such that $\xi^\alpha=f^{1/2}u^\alpha$ and at $f(\tilde{\rho})=1$ it normalizes to $\xi^\alpha \xi_\alpha=-1$. On the other hand, at the horizon $\xi^\alpha$ is null and both tangential and normal to the three dimensional horizon itself.  Thus, one can also identify this hyper-surface as the place where $\xi^\alpha\xi_\alpha=0$. Since $\xi^\alpha$ remains tangent to the null generator of the horizon it also satisfies the geodesic equation for some constant $\kappa$: 
  \be \label{sfgeod}
\xi_{\alpha;\beta}\xi^\beta=\kappa \xi_\alpha, 
\ee
which matches the $\kappa$ introduced in Eq.(\ref{sgfprime}) as we now review. First making use of the Killing equation, 
\be \label{killingeq}
\xi_{\alpha;\beta}+\xi_{\beta;\alpha}=0, 
\ee
equation \eqref{sfgeod} can be turned into:
\be
-\xi_{\beta;\alpha}\xi^\beta=\kappa \xi_\alpha,
\ee
or we can write it as 
\be \label{sfnormal}
-(\xi^\beta \xi_\beta)_{;\alpha}=2\kappa \xi_\alpha
\ee
which is not surprising since at the horizon $\xi_\alpha$ is normal to the $-(\xi^\beta \xi_\beta)$ constant surface. 
Now to calculate different sides of this equation and obtain $\kappa$ we use Eddington-Finkelstein ingoing or outgoing coordinates depending on which one is well suited to describe the horizon. For example, if the outgoing null ray is tangential to the horizon (as for the Schwarzschild metric) then the ingoing coordinate is appropriate:
\be 
ds^2=-f(\rho)dv^2+2dvd\rho+\rho^2d\Omega^2. 
\ee
However, if the ingoing null ray is tangential to the horizon (as for the static de-Sitter metric) then the outgoing coordinate is appropriate:
\be 
ds^2=-f(\rho)du^2-2dud\rho+\rho^2d\Omega^2. 
\ee
For Schwarzschild, $u$ constant and $\rho$ constant surfaces coincide at black hole horizon and are degenerate while for de Sitter $v$ constant and $\rho$ constant surfaces coincide.  
Transferring $\xi^\alpha$ to this coordinate and calculating $\xi_\alpha $ for these cases yields $\xi_0=-f$ and $\xi_1=\pm1$  where the plus and negative signs correspond  to $v$ and $u$ coordinates respectively. This implies that
\ba
-(\xi^\beta \xi_\beta)_{;\alpha}&=&f_{,\alpha}\nonumber\\&=& f^\prime \rho_{,\alpha}
\ea
  and also that at the limit where $f=0$, 
\be \label{xidown}
\xi_\alpha=\pm  \rho_{,\alpha} ~~~~~~(\text{at} ~\rho=\rho_H)
\ee
and together with Eq.(\ref{sfnormal}) we finally obtain:
\be
\kappa=\pm {1\over2 }f'(\rho_H).
\ee
It is easy to see intuitively why the signs here agree with $|f^\prime(\rho_H)|$ in Eq.(\ref{sgfprime}). The positive (negative) sign is for the case of outgoing (ingoing) null ray being tangential to the horizon, so it must be blocking the region $\rho<\rho_H$($\rho>\rho_H$) from observers. Therefore the observer $\tilde{\rho}$ must be at $\tilde{\rho}>\rho_H$ ($\tilde{\rho}<\rho_H$). Assuming $f(\rho)$ is monotonic and given that $f(\rho_H)=0$ and $f(\tilde{\rho})=1$, we expect $f^\prime(\rho_H)>0$ ($f^\prime(\rho_H)<0$). Notice that it is now evident why for these metrics defining horizon as $(\nabla \rho)^2=0$ is consistent with $ \xi^\beta \xi_\beta=0$. 
\subsection{Surface gravity in time dependent spherically symmetric spacetimes} \label{generaltrappinghorizon}
Let us start by writing a general spherically symmetric metric in the form
\be
ds^2= g_{ab}(r,t)dx^a dx^b+\rho^2(r,t)d\Omega^2 \,.
\ee
Here $g_{ab}$ is the metric for the $(t,r)$ coordinates and is sometimes referred to as the base metric and $\rho^2$ is the induced metric of two spheres at constant $r$ and $t$ and related to their areas through $\rho^2=A/4\pi$. A marginally trapped sphere or apparent horizon is defined such that at least one of  $\theta_+\equiv {d \ln A_+/ d\lambda}$ or $\theta_-\equiv {d \ln A_-/ d\lambda}$ vanishes along the null ray itself and consequently (as was explained in section \ref{defhorizon}) $(\nabla \rho)^2=0$ at apparent horizon. If $\theta_+$ and $\theta_-$ are zero at the same time, this is a bifurcation surface, but for the interest of this paper we consider the case in which only one of them is zero and it changes sign from one side to the other side. The union of all marginally trapped surfaces formed through extending the apparent horizon into past and future is called a trapping horizon and we will refer to it as $\Sigma$. Spheres for which $\theta_+\theta_-<0$, are called untrapped surfaces and one can fix the orientation of ingoing and outgoig in the region they reside by taking  $\theta_-<0$ and $\theta_+>0$~\cite{Hayward:1994bu}. On the other hand the region on the other side of $\Sigma$ is where trapped surfaces, defined by $\theta_+\theta_->0$, reside. To see how trapping horizons in cosmology differ from trapping horizons for blackhole and whitehole see table \ref{diffhorizons}. Next we consider cases where the apparent horizon is either a future outer trapping horizon or past inner trapping horizon and by that we mean: if $\theta_+=0$ ($\theta_-=0$) then $\theta_-\partial_- \theta_+>0$ ($\theta_+\partial_+ \theta_->0$). In the static limit this includes cases where apparent horizon is protecting the trapped side from sending light rays to untrapped side. 
\begin{table}[htdp]\label{diffhorizons}
\begin{center}
\resizebox {1\textwidth }{!}{ %
\begin{tabular}{|c|cccc|}
\hline
&Blackhole&Whitehole&Expanding FRW &Contracting FRW\\ 
Untrapped &$\theta_+>0$&$\theta_+>0$&$\theta_+>0$&$\theta_+>0$\\
 spheres&$\theta_-<0$&$\theta_-<0$&$\theta_-<0$&$\theta_-<0$\\
 &(outside)&(outside)&(inside)&(inside)\\
 \hline
 Trapped &$\theta_+<0$&$\theta_+>0$&$\theta_+>0$&$\theta_+<0$\\
 spheres&$\theta_-<0$&$\theta_->0$&$\theta_->0$&$\theta_-<0$\\
 &(inside)&(inside)&(outside)&(outside)\\
 \hline
Marginally &$\theta_+=0$ &$\theta_-=0$&$\theta_-=0$ &$\theta_+=0$\\
 trapped& $ \theta_-<0$(future)& $\theta_+>0$(past)& $ \theta_+>0$(past)&$ \theta_-<0$(future)\\
sphere&$\partial_-\theta_+<0$(outer)&$\partial_+\theta_-<0$(outer)&$\partial_+\theta_->0$(inner)&$\partial_-\theta_+>0$(inner)\\ \hline
\hline
\end{tabular}}
\end{center}
\caption{Characterizing trapping horizons based on expansion of congruence of light rays. The first row fixes the orientation of ingoing and outgoing vectors in untrapped region and removes the ambiguity for trapped region. $\partial_\pm$ denotes coordinate derivative along $l^a_\pm$.}
\end{table}%

As was discussed earlier, the definitions of $\kappa$ are intrinsically non-local and they also rely on the fact that the horizon is null and $\xi^\alpha$ is a Killing vector that satisfies the Killing equation \eqref{killingeq}. These criteria lead to  
\ba
\xi^\alpha(\xi_{\beta;\alpha}+\xi_{\alpha;\beta})&=&0 \label{contkilling}\\
\xi^\alpha(\xi_{\beta;\alpha}-\xi_{\alpha;\beta})&=&2\kappa\xi_\beta \label{antikodama}
\ea
and hence any linear combination of above equations can be used to define surface gravity as long as normalization of $\xi^\alpha$ is fixed. Note that also $\xi^\alpha$ becomes null at horizon and it is tangential to the null ray whose expansion vanishes. The Kodama vector $ \xi^a \equiv -\varepsilon^{ab}\nabla_b\rho$ is also null at marginally trapped surfaces and it is tangential to the null ray whose expansion vanishes (which we review in a few lines). Moreover, it goes into the Killing vector in the stationary limit of metric type Eq.(\ref{metricstatic}). It seems natural then to ask if substituting Kodama vector for Killing vector in the definition of surface gravity is possible. 

Unfortunately since $\xi^\alpha$ is no longer necessarily a Killing vector, the LHS of equation Eq.(\ref{antikodama})remains proportional to $\xi_\beta$ but the RHS of Eq.(\ref{contkilling}) is not necessarily zero and in general it is not even proportional to $\xi_\beta$. Even when $\Sigma$ is a null hyper-surface and the RHS of equation Eq.(\ref{contkilling}) is proportional to $\xi_\beta$, different linear combinations of the two equations result in different definitions for $\kappa$. However, as Hayward suggested \cite{Hayward:1997jp} since Eq.(\ref{antikodama}) remains valid for spherically symmetric metrics one can use that for defining surface gravity. Of course now the question of normalization remains since $\kappa$ clearly changes if $\xi$ is rescaled by a constant. How can we determine that the normalization of $\xi$ is consistent with normalization of Killing vector for null horizons? We review some proposed solutions in what follows. First, we show Kodama vector is in fact tangential to a appropriate null ray at $\Sigma$, second that its normalization is consistent with the local normalization of Killing vectors and last that this definition is consistent with 
\be
\kappa=\pm \frac{1}{2}\Box\rho  ~~~~~ \text{for}~~~\theta_\pm=0
\ee

We denote the outgoing and ingoing null vector as $l_+^\alpha$ and $l_-^\alpha$ respectively.  
Solving the second order null equation for this region yields:
\be \label{nullvectors}
 l_{\pm a} \propto g_{00} \delta^0_a + \left (g_{01}\mp \sqrt{-g}\right)\delta^1_a.
 \ee
 To determine which solution corresponds to which null  vector, $r$ constant surfaces were taken to be time-like (except at horizon itself) and $t$ constant surfaces to be space-like or null, thus $g_{00}$ must not be positive and $g_{11}$ must not be negative. On the other hand the condition $\theta_+=0$($\theta_-=0$) is equivalent to $l_{+}^a \nabla_a\rho=0$($l_{-}^a\nabla_a\rho=0$). Therefore, 
\ba \label{outinpropnull}
\nabla_a\rho &\propto &  l_{\pm a}     ~~~~~ \text{for}~~~\theta_\pm=0 
\ea
In fact, substituting Eq.(\ref{nullvectors}) in Eq.(\ref{outinpropnull}), the indices of Kodama vector can be lowered in a consistent way:
\ba \label{kodamalower}
\xi_a&=&\pm \nabla_a\rho  ~~~~~ \text{for}~~~\theta_\pm=0 \\
&\propto& l_{\pm a}  ~~~~~ \text{for}~~~\theta_\pm=0 
\ea
As we discussed before, even though Eq.(\ref{sfgeod}) is a local equation that can be applied to any Killing horizon it does not provide a local definition of surface gravity, since $\xi$ was not normalized locally, independent of $\tilde{\rho}$. To overcome this challenge an alternative method has been put forward \cite{Ashtekar:1999yj} which relies on first normalizing the auxiliary null vector field $N^a$, defined by $N^a\xi_a=-1$. In fact this method can also be used to fix the rescaling of the Kodama vector. 
The prescription is that one can impose the normalization by fixing the nonzero expansion of null rays at $\Sigma$ itself:
\be
\theta _\mp= \mp{2 \over \rho} ~~~~~\text{for}~~~\theta_\pm=0 
\ee
For example let us  rescale $\xi_\alpha$ by a constant $\xi_\alpha \rightarrow \tilde{\xi}_{\alpha}=\pm C~ \nabla_{\alpha}\rho$. Associating  $N^\alpha$ to the other null beam tangent to the null geodesic, $N^\alpha= {dx^\alpha\over d\lambda}|_{N}$ ,  the above condition leads to: 
\ba
 \theta_\mp&=&{2 \over \rho} \rho_{,a} N^\alpha \nonumber\\
 &=&\mp{2 \over \rho}~.
\ea
Thus $\rho_{,a} N^\alpha=\mp 1$ and further fixing of $N^a\xi_a=-1$ immediately results in $C=1$. Now that $\xi$ is normalized it is justified to use Equation \eqref{antikodama} to calculate $\kappa$.  This equation can be expressed as
\be  \label{antikodama2}
\xi^a\nabla_{[a}\xi_{b]}=\kappa \xi_b\;.
\ee
To prove this equation note that $\nabla_{[a}\xi_{b]}=[ab] \nabla_{[0}\xi_{1]}$, where $[ab]$ represents the permutation symbol. This implies 
\ba
\xi^a\nabla_{[a}\xi_{b]}&=&-[ab]\nabla_{[0}\xi_{1]}\epsilon^{ac}\nabla_c\rho \nonumber\\
&=&{\nabla_{[0}\xi_{1]}\over\sqrt{-g}} \nabla_b \rho\nonumber\\
&=&(\pm {\nabla_{[0}\xi_{1]}\over\sqrt{-g}}) \xi_b ~,
\ea
where to arrive at the last line we used Eq.(\ref{kodamalower}). We now see that Equations \eqref{antikodama} and \eqref{antikodama2} hold at $\Sigma$ if 
\be
\kappa\equiv \pm {\nabla_{[0}\xi_{1]}\over\sqrt{-g}}  ~~~~~\text{for}~~~\theta_\pm=0~.
\ee
Substituting for $\xi_a=g_{ac}\xi^c$ and using the definition of Kodama vector, we observe that  
\ba
2 \nabla_{[0}\xi_{1]} &=& [c d] \nabla_c \xi_d \\
&=& \sqrt{-g} \epsilon^{cd}\nabla_c(g_{d e} \epsilon^{e f} \nabla_f \rho)\\
&=&  \sqrt{-g} \Box \rho ,
\ea
where in the last line we have used $ \epsilon^{cd} \epsilon^{e f} g_{d e} =  g^{cf}$ and that  Levi-Civiata tensor is covariantly constant with respect to $(1+1)$ covariant derivative.
Therefore
\be
\kappa=\pm \frac{1}{2}\Box\rho ~~~~~\text{for}~~~\theta_\pm=0~.
\ee
It is worth noting that, for a slow-roll inflationary back-ground, the definition put forward in \cite{2004PhRvL..92a1102B} (based on projected acceleration of the vector tangential to the trapped surface) differs from the above at $O(\epsilon)$.

\bibliographystyle{JHEP}
\bibliography{BH}

\end{document}